\documentclass[final,5p,times,twocolumn]{elsarticle}

\usepackage{amsmath}
\usepackage{amssymb}
\usepackage{amsfonts}
\usepackage{epsfig}

\usepackage{natbib}

\biboptions{square,sort&compress}

\def\dd{{\rm d}}

\def\L{{\cal L}}

\def\P{{\cal P}}
\def\U{{\cal U}}

\def\Veff{V_{\text{eff}}}

\def\T{\mathcal{T}}
\def\abs#1{\left|#1\right|}

\begin{document}

\begin{frontmatter}

\title{A consistent scalar-tensor cosmology for inflation, dark energy and the Hubble parameter}

\author[ab,ral]{C. H.-T. Wang}
\ead{c.wang@abdn.ac.uk}

\author[ab]{J. A. Reid}

\author[ed]{A. St. J. Murphy}

\author[ab]{D. Rodrigues}

\author[ab]{M. Al Alawi}

\author[ral,st]{R. Bingham}

\author[ist]{J. T. Mendon\c{c}a}

\author[ab]{T. B. Davies}

\address[ab]{Department of Physics, University of Aberdeen, King's College, Aberdeen AB24 3UE, UK}

\address[ral]{Rutherford Appleton Laboratory, Chilton, Didcot, Oxfordshire OX11 0QX, UK}

\address[ed]{School of Physics and Astronomy, University of Edinburgh, Edinburgh, EH9 3JZ, UK}

\address[st]{Department of Physics, University of Strathclyde, Glasgow G4 0NG, UK}

\address[ist]{IPFN, Instituto Superior T\'ecnico, 1049-001 Lisboa, Portugal}


\begin{abstract}
A Friedman cosmology is investigated based on scalar-tensor gravitation with general metric coupling and scalar potential functions. We show that for a broad class of such functions, the scalar field can be dynamically trapped using a recently suggested mechanism. The trapped scalar can drive inflation and accelerated cosmic expansion, compatible with standard requirements. The inflationary phase admits a natural exit with a value of the Hubble parameter dictated by the duration of inflation in a parameter independent manner. For inflationary duration consistent with the GUT description, the resulting Hubble parameter is found to be consistent with its observed value.
\end{abstract}

\begin{keyword}
Modified theories of gravity\sep
cosmic inflation\sep
dark energy
\PACS 98.80.Cq, 04.50.Kd, 95.36.+x
\end{keyword}

\end{frontmatter}

\section{Key hypotheses, observations and their context}
\label{sec:intro}

Rapid inflation after the Big Bang, slow expansion of the current universe with a recurring acceleration, and core-collapse supernovae are extraordinary gravitational events in modern astronomy and cosmology but have not been discussed under a unified theoretical framework. Given the uncertain physical mechanisms involved in these events to various degrees, it is perhaps counterintuitive to attempt a unified theoretical approach. Surprisingly, however, using a wide class of simple extensions to Einstein's theory of gravity, general relativity (GR), a mechanism appears to exist under which key features of these distinct events could follow as a dynamical consequence.

Gravitational interaction with a scalar field is an indispensable ingredient of most theories of inflation in the early universe with a potential that allows for either a false vacuum \cite{Guth1981} or a slow roll \cite{Linde1982} dynamical behavior of the scalar. There are fundamental reasons for a scalar field to be in fact an intrinsic part of gravitational theory from the viewpoints of the equivalence principle, quantum gravity, and unified theory and its low energy reduction, leading to a variety of scalar-tensor (ST) theories of gravity \cite{Brans1961, Dicke1962, Damour1993b, varaoni, Khoury2004, Wang2011, Wagoner1970}, together with intense effort of experimental tests \cite{Bertotti2003, Damour1998, Scharre2002,  Steffen2010, Margerin2014}. Furthermore, there has been substantial interest in incorporating ST theory into (extended) inflation \cite{Steinhardt1990, LaSteinhardt1989} leading to significant developments as summarized in \cite{Clifton2012}.

In a recent paper \cite{Wang2013}, a new type of trapping of the scalar gravitational field for a broad class of ST theories has been identified. Remarkably, it allows the scalar gravitational field to not only to drive inflation which could be tested through cosmological imprints such as CMB anisotropies, but also to potentially reinvigorate core-collapse supernova
by topping up apparently missing energy required to explain the observed powerful astronomical explosions.

Our scalar trapping mechanism involves an effective scalar potential activated dynamically through the matter stress tensor, metric coupling and scalar potential, in a fashion analogous to the ``chameleon effect'' \cite{Khoury2004} that has received  major attention in the recent literature \cite{Clifton2012}.
However, our ST formulation has distinct properties to be made precise below in terms of a set of conditions on the metric coupling and scalar potential. In particular, the trapped scalar gravitational field to be discussed here can have a large enough value to generate important effects on violent gravitational events. Here we show that such a dynamical trapping mechanism, when applied to the standard homogeneous and isotropic cosmological model, may indeed yield a generic inflationary behavior that allows for realistic e-folding and duration.

Remarkably, the model admits a natural exit from inflation after a duration, which if chosen to be consistent with the grand unification theory (GUT) description of the early universe, yields an exit Hubble parameter value eventually evolving into the measured present day value. The statement is largely parameter independent and requires only an effective trapping of the scalar field during inflation that can be satisfied by a large class of metric coupling and scalar potential functions. We then show that the applicability of the model is not limited to the high energy domain and can also give rise to an emergent dark energy in the late universe consistent with the observational value.


\section{Cosmic evolution of metric and scalar fields}
\label{sec:ST}

Under the metric description of gravitation compatible with the equivalence principle, ST theory is a natural extension to GR, by including a scalar field $\phi$ as part of gravity. It dynamically rescales the traditional ``Einstein'' metric tensor  $g_{ab}$ into the ``physical'' metric tensor
$
\bar{g}_{ab} = \Omega(\phi)^2g_{ab}
$
which interacts directly with matter, using a theory-dependent positive metric coupling function $\Omega(\phi)$.


However, we will use $g_{ab}$ as the dynamical metric in the following analysis since the corresponding generalized Einstein and Friedman equations offer a direct connection with the conventional ones. Furthermore, the scalar interaction in the Einstein frame can be interpreted as mass rescaling using $\phi$ together with a geometrical description using $g_{ab}$ \cite{Wagoner1970, Bekenstein1980}.
We choose the metric signature $(-,+,+,+)$ with spacetime coordinate indices $a,b,\cdots=0,1,2,3$, denote by $\phi_0$ the current cosmological value of $\phi$, and adopt the convenient convention $\Omega(\phi_0)=1$ and $\phi_0 =0$.

The general field equations
\begin{eqnarray}
G_{ab} = \frac{8\pi G}{c^4}\, (\T_{ab} + T_{ab}{})
\label{geq}
\end{eqnarray}
\begin{eqnarray}
\Box\, \phi - V'(\phi) + \frac{4\pi G}{c^4}\,  A(\phi)T{} = 0 .
\label{phieq}
\end{eqnarray}
in the Einstein frame of ST theory for the dimensionless metric tensor $g_{ab}$ and scalar field $\phi$ are derived from the Lagrangian for ST theory of gravity interacting with matter given by  \cite{Wagoner1970, Wang2013}:
\begin{eqnarray}
\frac{c^4}{16\pi G} \int{\dd^3x} \,g^{1/2}R^{(g)} + \L + L{}
\label{act}
\end{eqnarray}
where
\begin{eqnarray}
\L = -\frac{c^4}{4\pi G} \int{\dd^3x} \,g^{1/2}
\left[\frac12\,g^{ab}\phi_{,a}\phi_{,b} + V(\phi)\right]
\label{actphi}
\end{eqnarray}
and $L{}$ are the Lagrangians for the scalar field $\phi$ and matter respectively.

Here $c$ is the speed of light, $G$ is the gravitational constant, $\Box$ is the Laplace-Beltrami operator, $G_{ab}$ is the Einstein tensor, $T_{ab}$ and $\T_{ab}$ are matter and scalar effective stress tensors respectively, $T$ is the contracted matter stress tensor with respect to $g_{ab}$, $V(\phi)$ is a theory dependent scalar potential, and
$ A(\phi)=(\ln \Omega(\phi))'$
with a prime denoting differentiation with respect to $\phi$.

Eq. \eqref{geq} is the generalized Einstein equation and Eq. \eqref{phieq} is the scalar field equation, which in the following will be analyzed in the standard Robertson-Walker cosmology with the scale factor $a(t)$, Hubble parameter $H(t)={\dot{a}}/{a}$ denoting time derivative by an over-dot.
The curvature parameter has been neglected in Eq. \eqref{phieq} as its effects are known to be quickly diminished during inflation \cite{Carroll2001}.

In terms of the energy density $u$ and pressure $p$ of matter and effective energy density
\begin{eqnarray}
\U
=
\frac{c^2}{8\pi G} \left[
\dot\phi^2 + 2c^2V(\phi)
\right]
\label{U}
\end{eqnarray}
and effective pressure
\begin{eqnarray}
\P
=
\frac{c^2}{8\pi G} \left[
\dot\phi^2 - 2c^2V(\phi)
\right]
\label{P}
\end{eqnarray}
of the scalar field $\phi$, the effective Einstein equation \eqref{geq} yields the Friedmann equations
\begin{eqnarray}
H^2
=
\frac{8\pi G}{3c^2}\, (\U + u)
\label{F1}
\end{eqnarray}
and
\begin{eqnarray}
\frac{\ddot{a}}{a}
=
-\frac{4\pi G}{3c^2}
\left(
\U + 3\P + u + 3p
\right) .
\label{F2}
\end{eqnarray}

As is widely used, here the matter content will be represented with a cosmological fluid with the equation of state $p=w u$ with a constant pressure to energy density ratio $w$ in the normal range $-1/3 < w < 1/3$ implying inflation can only be driven by the scalar field. Then by using Eqs. \eqref{F1}  \eqref{F1}  and \eqref{F1}, we see that Eqs. \eqref{geq} and \eqref{phieq} become respectively the following Friedmann and scalar equations:
\begin{eqnarray}
&&\hspace{-20pt}
\dot{H}
+\frac32(1+w)H^2
+\frac12(1-w)\dot\phi^2
-c^2(1+w)V
=0
\label{Eq1}
\\[3pt]
&&\hspace{-20pt}
\ddot\phi
+3H\dot\phi
-
\frac{1}{2}
(1-3w)\,  A
\dot\phi^2
\nonumber\\[3pt]
&&
\hspace{18pt}
+
(1-3w)\Big(
\frac32H^2
-
c^2
V\Big) A
+
c^2V'
= 0.
\label{Eq2}
\end{eqnarray}
As with the scalar potential used for slow roll inflation, we will consider $V(\phi)$ to be a monotonic ascending positive function:
\begin{eqnarray}
V(\phi) > 0
,\quad
V'(\phi) > 0
\label{VV}
\end{eqnarray}
together with the well-known slow roll conditions \cite{Carroll2001}:
\begin{eqnarray}
\dot\phi^2 < c^2V
,\quad
\ddot\phi < \frac12\, c^2V' .
\label{slowroll}
\end{eqnarray}
From Eqs. \eqref{U} and \eqref{F1}, a useful property directly associated with the positivity of the matter density $u>0$ is
\begin{eqnarray}
\frac32H^2 - c^2V
>
\frac{1}{2}
\dot\phi^2 .
\label{FF1a}
\end{eqnarray}
Since the right-hand side (RHS) of \eqref{FF1a} is positive, so must be its left-hand side (LHS). It follows that the second last term on the LHS of Eq. \eqref{Eq2} is negative provided that
\begin{eqnarray}
A(\phi) < 0.
\label{AA}
\end{eqnarray}
In this case, that term effectively acts as an opposing force to that of the last term on the  LHS of \eqref{Eq2}. Eq. \eqref{AA} means, in comparison with the properties of the scalar potential in \eqref{VV}, the metric coupling function is a monotonic descending positive function:
\begin{eqnarray}
\Omega(\phi) > 0
,\quad
\Omega'(\phi) < 0 .
\label{OO}
\end{eqnarray}

Consequently, together with \eqref{slowroll}, the downhill rolling of $\phi$ could so slowly track the local minimum of an effective trapping potential satisfying
\begin{eqnarray}
\Veff'(\phi) := V'(\phi) - \frac{4\pi G}{c^4}\,  A(\phi)T{} = 0
\label{phieq1}
\end{eqnarray}
obtained from setting $\Box\, \phi=0$ in \eqref{phieq}, that all the time derivatives of $\phi$ in \eqref{Eq1} and \eqref{Eq2} may now be treated as negligible. Therefore, these two equations reduce to
\begin{eqnarray}
(1/H)\,\dot{}
=
q+1
=
\frac{3(1+w)}
{
2
-
2(1-3w) A
{V}/
{V'}}
\label{Eq1f}
\end{eqnarray}
\begin{eqnarray}
H^2
=
-\frac{2c^2V'}{3(1-3w) A}
+
\frac23
c^2
V
\label{HV}
\end{eqnarray}
where $q$ is the deceleration parameter.

\section{Inflation driven by trapped scalar}
\label{sec:infl}

Eqs. \eqref{Eq1f} and \eqref{HV} form a closed differential-algebraic system for $H$ and $\phi$. This system can be solved first by regarding $q+1$ as an expression of $\phi$ given by the RHS of \eqref{Eq1f}. In turn, $\phi$ is implicitly and yet uniquely determined by a positive $H$  through Eq.~\eqref{HV} if its RHS is a monotonic ascending function of $\phi$, satisfying
\begin{eqnarray}
A V''-A'V' < 0
\label{AV0}
\end{eqnarray}
obtained from $H' > 0$ using Eqs. \eqref{VV} and \eqref{Eq1f}.
In this way, as with the slow roll inflation, both $H$ and $\phi$ could decrease until the end of inflation.
At the same time, $q$ must be a monotonic descending function of $\phi$ so that it can increase with decreasing $H$ during inflation until $q=0$. This requirement can be cast into the following inequality:
\begin{eqnarray}
A V'{}^2 < V(A V''-A'V') .
\label{AV1}
\end{eqnarray}
Conditions \eqref{VV}, \eqref{OO}, \eqref{AV0} and \eqref{AV1} can be satisfied by a large class of metric coupling and potential functions, for example:
\begin{eqnarray}
\Omega(\phi)
&=&
e^{\frac12\beta \phi^2} \Rightarrow A(\phi) = \beta\phi,
\label{OA}
\\[3pt]
V(\phi)
&=&
\xi\phi^\sigma
\label{OV}
\end{eqnarray}
for $\phi>0$ with $\beta < 0$, $\xi > 0$ and $\sigma > 2$.

On applying these general conditions to Eq. \eqref{Eq1f}, we see that during inflation the deceleration parameter increases in the following range:
\begin{eqnarray}
-1 < q < 0
\label{Q}
\end{eqnarray}
corresponding to the decreasing of the LHS of the relation:
\begin{eqnarray}
-\frac{2 A V}{V'} \ge {\frac{1+3w}{1-3w}}
\label{Eq1d}
\end{eqnarray}
until its equal sign holds with $q=0$, which defines the time $t_f$ and the value of the scalar $\phi_f = \phi(t_f)$ at the end of inflation. We will similarly use the subscript $f$ to denote the other quantities at the end of inflation.

Using the argument leading to the conditions \eqref{AV1}, showing the LHS of \eqref{Eq1d} decreases with decreasing $\phi$, we see that inflation lasts while $\phi > \phi_f$  with the corresponding exit value $H_f$ evaluated through \eqref{HV}. However, it is important to note that, as $w\to1/3$ (for a radiation fluid) we have $\phi_f\to\infty$ and so inflation cannot occur. On the other hand, as $w\to-1/3$ (for a strong energy condition violating fluid) we have $\phi_f\to 0$ and so inflation goes on for any $\phi > 0$.

Provided that the above conditions on $A(\phi)$ and $V(\phi)$ are satisfied, we may now proceed to estimate quantities associated with inflation. Denote by $t_i$ as the initial time at the start of inflation and $H_i = H(t_i)$. Denoting by $\tau= {t_f-t_i}$ the duration and by $N = \ln a(t_f) - \ln a(t_i)$ the e-folding of the inflation, we see from \eqref{Eq1f} and \eqref{Q} that
\begin{eqnarray}
\frac1{H_f} - \frac1{H_i} \lesssim \tau
\label{dur}
\end{eqnarray}
\begin{eqnarray}
{H_f}\, \tau < N < {H_i}\, \tau.
\label{ef}
\end{eqnarray}

From \eqref{ef} a sufficiently large e-folding $N$ required to address primordial nucleosynthesis, flatness and horizon problems in cosmology may be obtained if $H_i \gg H_f$, under which an order of magnitude estimate yields $H_f \approx 1/\tau$. In terms of the Planck time $t_P$, scenarios of inflation generally suggest it starts at the end of the grand unification epoch with $t_i \approx 10^7 t_P$ and finishes before the electroweak epoch with $t_f \approx 10^{10} t_P$  after the Big Bang (Ref. \cite{Planck2013} and references therein.)
Using these values we see that ${H_f} \approx 10^{-10} /t_P$. Together with the standard assumption that after the end of inflation the matter dominant $H \propto 1/t$ lasts for $t_0 = 10^{61} t_P$ so that we obtain $H_0 = (t_f/t_0) H_f \approx 10^{-61} /t_P$, in agreement with the observed value of the present day Hubble parameter.

In addition, the initial Hubble parameter may now be chosen more naturally to be closer to the Planck scale of $H_i \sim 1/t_P$ leading to a very large upper bound of the e-folding $N$ up to $10^{10}$. By contrast, exponential inflation with a constant Hubble parameter $H_i = H_f \approx 10^{-10} /t_P$ would require a much longer duration in excess of $\tau \approx  10^{12} t_P$ for a minimum e-folding $N$ of $60$.

\section{Trapped scalar as dark energy}
\label{sec:DE}

A striking feature of the ST cosmology being discussed here is that the described universe can undergo multiple accelerating or decelerating phases, even though the scalar potential $V(\phi)$ and metric coupling $\Omega(\phi)$ are monotonic functions overall according to Eqs. \eqref{VV} and \eqref{OO}. For even after the inflationary era with continuously decreasing Hubble parameter and scalar field, as long as the conditions for acceleration discussed above are met, in particular Eq. \eqref{Eq1d}, the expansion of the universe is ready to return to such a phase again. This possible resurrection of accelerating expansion is clearly relevant for the ongoing dark energy problem \cite{Carroll2001, Peebles2003, Planck2013a}.

Recall that we choose the cosmological value of the scalar $\phi_0 = 0$ as the reference value for the scalar field and so to be sure, condition \eqref{AV1} is only required for $\phi \gg 0$ to allow for inflation in the early universe. To test the candidacy of $\phi$ as (part of) dark energy, we shall explore the property of $\Omega(\phi)$ and $V(\phi)$ in the current era with a smaller value of ${\phi}$, with the understanding that they continue into their function values monotonically according to \eqref{VV} and \eqref{OO} for larger $\phi$ values that may count for inflation. For the cosmic expansion to return to acceleration, $q$ should now decrease from a positive value into the range given by \eqref{Q}. Reversing the argument leading to \eqref{AV1}, we now arrive at the condition
\begin{eqnarray}
A V'{}^2 > V(A V''-A'V')
\label{AV2a}
\end{eqnarray}
for the metric coupling and scalar potential to be satisfied for smaller $\phi$ in the late universe, while still holding Eq. \eqref{AV0}.

In general, given an observed or modelled Hubble parameter $H(t)$ for any stage of the universe, we can infer from \eqref{Eq1f} and \eqref{HV} that
\begin{eqnarray}
A(t)
&=&
\frac{3(1+w)H(t)}{1-3w}
+
\frac{\ddot{H}(t)}{(1-3w)\dot{H}(t)}
\label{Aeq}
\\
V(t)
&=&
\frac{3H^2(t)}{2c^2}
+
\frac{\dot{H}(t)}{c^2(1+w)} .
\label{Veq}
\end{eqnarray}
If the form of $A(\phi)$ is known as a monotonic function and hence its inverse $\phi(A)$ as well then together with \eqref{Aeq} we have
\begin{eqnarray}
\phi(t)=\phi(A(t))
\label{phiteq}
\end{eqnarray}
allowing us to recover the scalar potential $V(\phi)$ by using
\begin{eqnarray}
V(\phi)=V(t(\phi)) .
\label{Vphit}
\end{eqnarray}
Alternatively, if the form of $V(\phi)$ is known as a monotonic function and hence its inverse $\phi(V)$ as well then together with \eqref{Veq} we have
\begin{eqnarray}
\phi(t)=\phi(V(t))
\label{phiteqV}
\end{eqnarray}
allowing us to recover  $A(\phi)$ by using
\begin{eqnarray}
A(\phi)=A(t(\phi)) .
\label{Aphit}
\end{eqnarray}

Using the $\Lambda$CDM model \cite{Mukhanov2005} for the universe dominated by matter density ratio $\Omega_m$ and effective cosmological constant for dark energy density ratio $\Omega_\Lambda$ so that
\begin{eqnarray}
\Omega_m+\Omega_\Lambda=1
\end{eqnarray}
with the cosmological constant \cite{Barrow2011}
\begin{eqnarray}
\Lambda
=
\frac{3H_0^2\Omega_\Lambda}{c^2}
\label{Lamb}
\end{eqnarray}
the standard Friedmann equation reads
\begin{eqnarray}
\dot{H}^2(t)
=
H_0^2
\left(
\frac{\Omega_m}{a^3(t)}+\Omega_\Lambda
\right)
\label{Feq}
\end{eqnarray}
having the solution
\begin{eqnarray}
H(t)
=
H_0\sqrt{\Omega_\Lambda}\coth
\frac{3H_0\sqrt{\Omega_\Lambda}\:t}{2}
\label{HHeq}
\end{eqnarray}
using the convention $a(t_0)=1$.

Substituting \eqref{HHeq} into \eqref{Aeq} and \eqref{Veq}, we immediately obtain
\begin{eqnarray}
A(\phi)
&=&
0
\label{Aeq2}
\\
V(\phi)
&=&
\frac{\Lambda}{2}
\label{Veq2}
\end{eqnarray}
for the ST description of the universe dominated by matter and dark energy with associated cosmological constant $\Lambda$ given by \eqref{Lamb}. Therefore the accelerating cosmic expansion consistent with the $\Lambda$CDM model is recovered with the scalar-metric coupling and scalar potential tending to \eqref{Aeq2} and \eqref{Veq2} for small $\phi$ in the late universe.

It has been expected that the original and simplest ST theory due to Brans and Dicke \cite{Brans1961,Dicke1962} with the coupling function
\begin{eqnarray}
\Omega(\phi) = e^{\alpha\phi} \Rightarrow A(\phi) = \alpha
\label{BD}
\end{eqnarray}
where $\alpha$ is a constant and is related to the Brans-Dicke parameter $\omega$ by $\alpha^2=(2\omega+3)^{-1}$, provides a good approximation of a more general ST theory in the low energy domain with a small $\phi$ for the present epoch of the universe \cite{Wagoner1970}. The current constraint on the Brans-Dicke parameter is $\omega > 40000$ \cite{Bertotti2003}, placing $\abs{\alpha} < 0.004$, which is consistent with our condition \eqref{Aeq2} for the scalar-metric coupling derived from general relations \eqref{Aeq} and \eqref{Veq} applied to the late universe.

The ST cosmology discussed here uses a scalar potential with a value expected to decrease continuously from the early to late universe while the emerging cosmological constant becomes active in the very early and very late universe separately.
Additionally, it is important to note that, our scalar potential could theoretically continue to decrease into the future universe resulting in deviation from the $\Lambda$CDM model.

\section{Overall cosmic evolution and discussions}
\begin{figure}
\begin{center}
\includegraphics[width=0.4\linewidth]{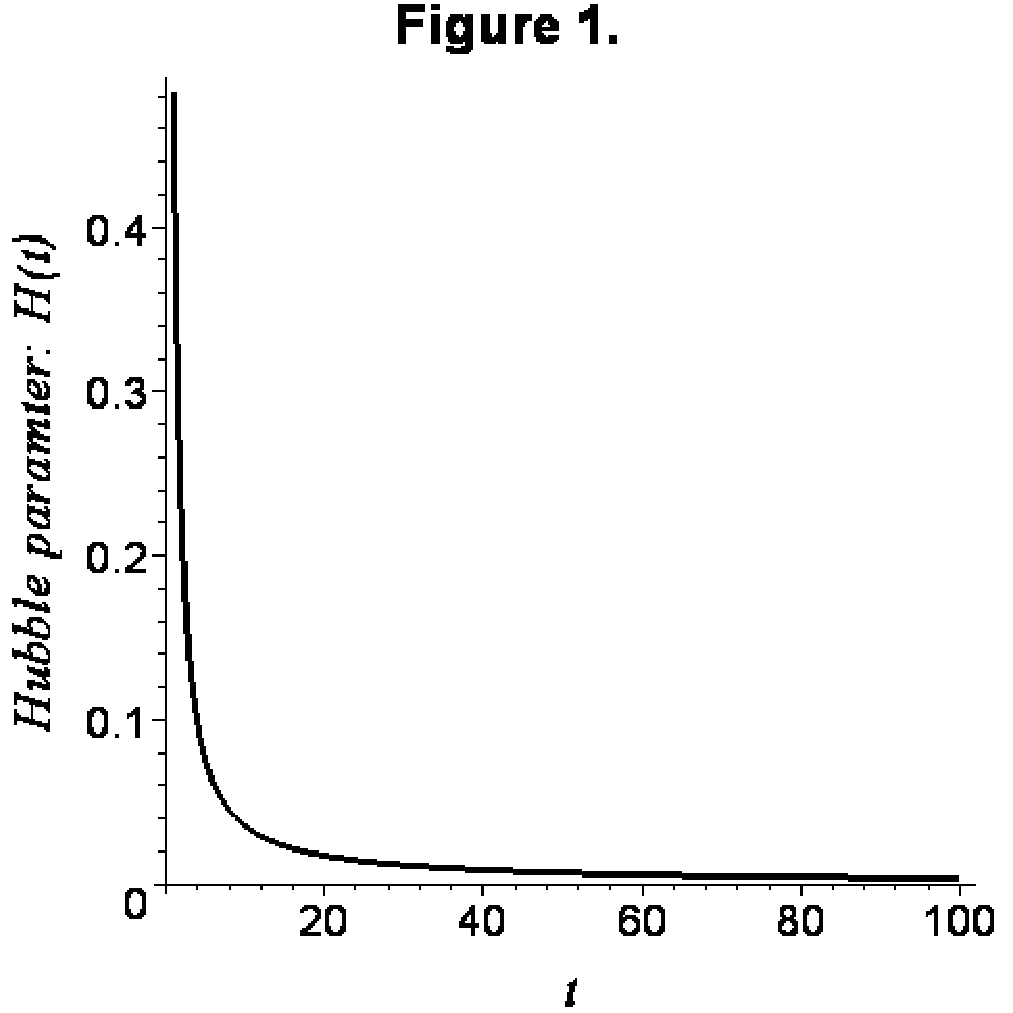}
~~
\includegraphics[width=0.4\linewidth]{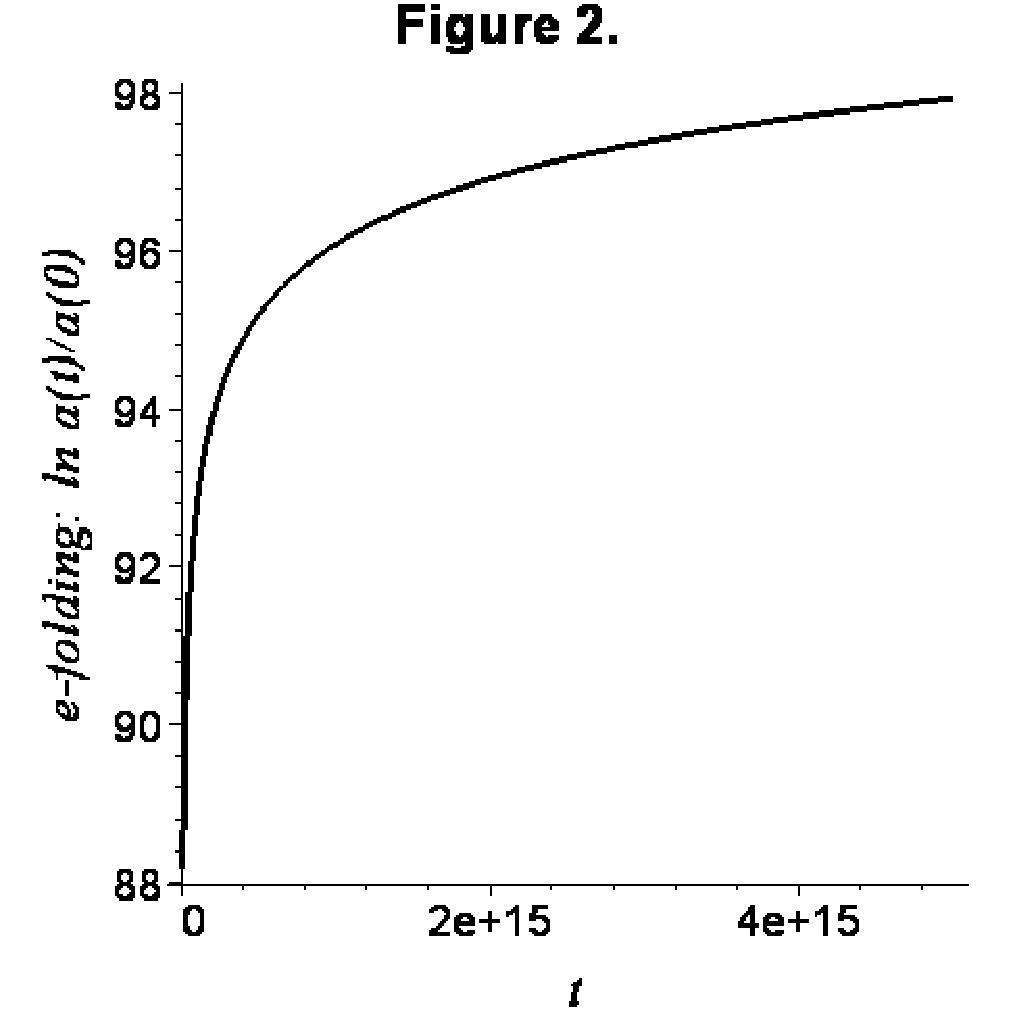}

\vskip 15pt

\includegraphics[width=0.4\linewidth]{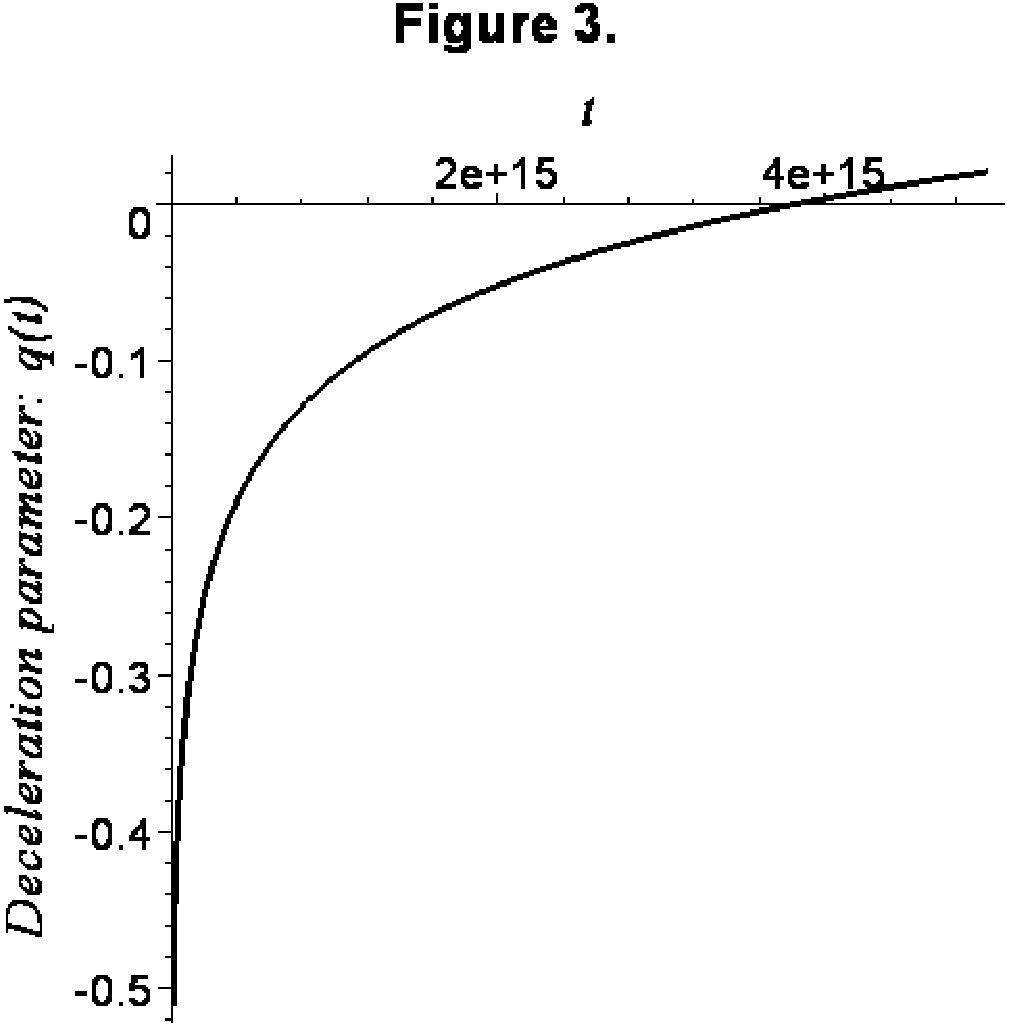}
~~
\includegraphics[width=0.4\linewidth]{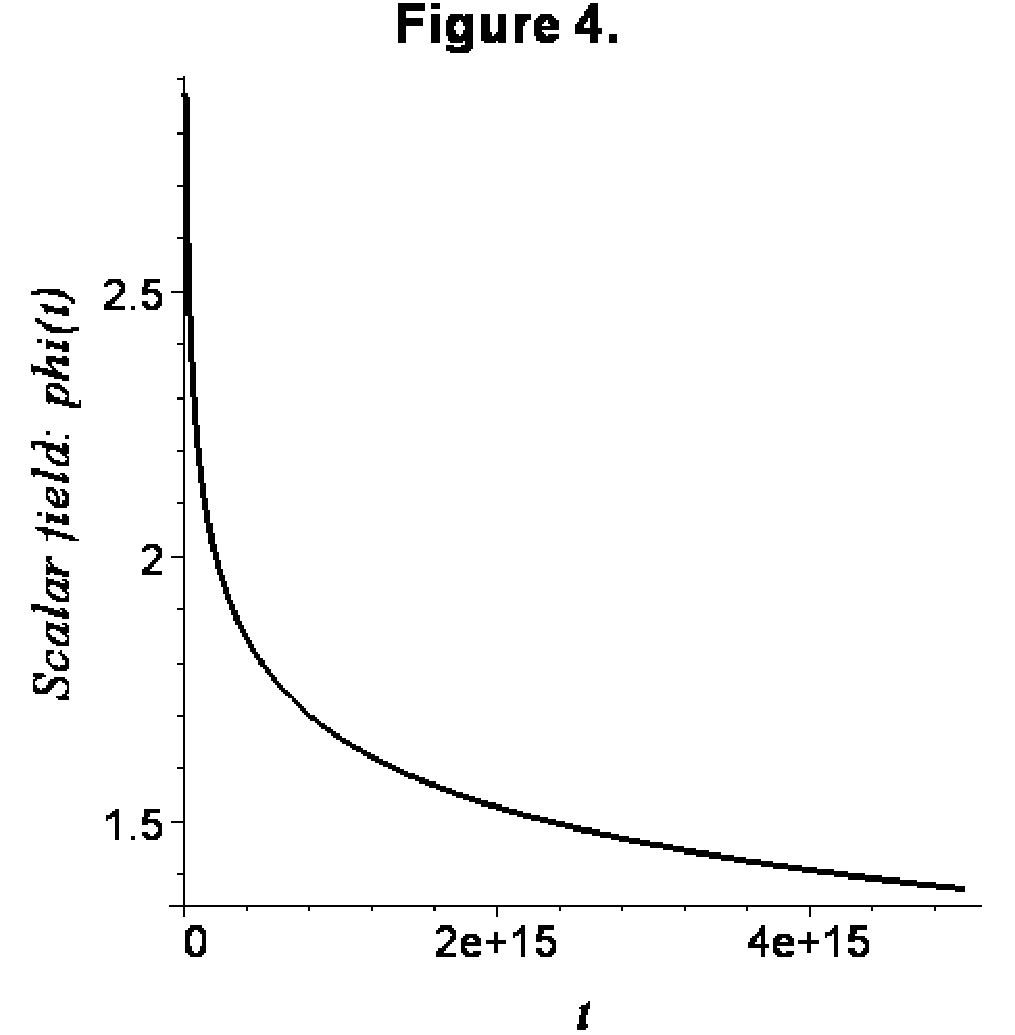}
\end{center}

\vskip 10pt

\begin{small}
{\bf Figures 1--4:}
To demonstrate an overall cosmic evolution, direct numerical integrations of  Friedmann  and  scalar equations \eqref{Eq1} and \eqref{Eq2} along with metric coupling \eqref{OA2} and scalar potential \eqref{OV2} have been carried out in the Planck units where $G = c = \hbar = 1$. As exemplified above with illustrative values $\beta=-5, \sigma=20, \xi=4\times10^{-35}, w=0$, we obtain a possible evolution of the Hubble parameter in Fig. 1, which has e-folding $>60$ during inflation shown in Fig. 2. This inflation ends automatically at $t\sim 4\times10^{15}$ when the model universe enters into a matter dominated decelerating era. Meanwhile, Fig. 4 shows the trapped scalar field continues to fall off so that the metric coupling \eqref{OA2} and scalar potential \eqref{OV2} will eventually reduce to  \eqref{Aeq2} and \eqref{Veq2} describing the $\Lambda$CDM model where the accelerated cosmic expansion will return at a moderate pace with an emergent dark energy phenomenon.
\end{small}
\end{figure}

Following our treatments of inflation in section \ref{sec:infl} and dark energy in section \ref{sec:DE} using the same ST framework for cosmology described in section \ref{sec:ST}, we would now like to consider an integrated approach. Its possibility is significant since inflation and dark energy in the early and late universe respectively have in many ways distinct phenomenological bases and subtleties. The sufficient e-folding for satisfactory inflation is not only required for flatness and horizon problems described in section \ref{sec:intro}, but also necessary for the amplification of the quantum fluctuations of the very early universe to allow for structure formations that seed stars and galaxies \cite{Liddle2000, Liddle2003, Lyth2009}. This in particular results in the spectrum of perturbations to be nearly scale-invariant \cite{Zeldovich1972, Harrison1970, Peebles1970} as have been observationally confirmed by the WMAP mission and other cosmic microwave background surveys \cite{WMAP2013}. In the present work, we focus on the possibility of generating sufficient inflation necessary for such a process using a unified ST cosmology framework leaving more detailed perturbation analysis for future investigations. The dark energy on the other hand, has been necessitated by the observed acceleration of the cosmic expansion through type Ia supernovae redshift measurements \cite{Riess1998, Perlmutter1999, Riess2001, Durrer2011}, as well as further confirmation by the WMAP \cite{WMAP2013} and Planck  \cite{Planck2013a, Planck2013xvi} missions, resulting in the $\Lambda$CDM model described in section \ref{sec:DE}. Within our current ST cosmology formalism, it is interesting to notice that the metric couplings and scalar potentials featured in section \ref{sec:infl}, namely \eqref{OA} and \eqref{OV}, and in section \ref{sec:DE}, namely \eqref{Aeq2} and \eqref{Veq2} may arise as limiting cases from more general expressions such as:
\begin{eqnarray}
A(\phi)
&=&
\beta \phi \, \theta(\phi)
\label{OA2}
\\[3pt]
V(\phi)
&=&
\xi\phi^\sigma\, \theta(\phi) + \frac{\Lambda}{2}
\label{OV2}
\end{eqnarray}
where $\theta$ denotes the Heaviside step function, for large and small $\phi$ in the early and late epochs of the universe respectively. This is illustrated with the numerical integrations of the resulting  Friedmann  and  scalar equations \eqref{Eq1} and \eqref{Eq2} described in figures 1--4 that recover features of inflation and dark energy in a single model.
Following our reported work, further details and refinements of discussed novel features of the related ST cosmology and their wider implications may deserve to be explored theoretically and experimentally.

\section*{Acknowledgments}

The authors are grateful for financial support to
the Cruickshank Trust (CW),
EPSRC/GG-Top (CW, JR),
Omani Government (MA),
Science Without Borders programme, CNPq, Brazil (DR),
and
STFC/CfFP (CW, AM, RB, JM).
CW and AM acknowledge the hospitality of CERN, where this work was started.
The University of Aberdeen and University of Edinburgh are charitable bodies registered in Scotland, with respective registration numbers SC013683 and SC005336.


\end{document}